\documentclass[12pt]{article}

\usepackage{epsf}
\usepackage{amsmath}
\usepackage{cite}
\usepackage{graphics}
\usepackage{amsfonts}
\usepackage{amssymb}
\usepackage{latexsym}
\usepackage{color}
\input{colordvi.tex}

\renewcommand{\thefootnote}{\#\arabic{footnote}}

\begin{document}
\setcounter{footnote}{0}

\begin{titlepage}

\begin{center}

\hfill July 2008\\

\vskip .5in

{\Large \bf
Signatures of Non-Gaussianity in the Curvaton Model

}

\vskip .45in

{\large
Kari Enqvist$\,^{1,2}$ and Tomo Takahashi$\,^3$
}

\vskip .45in

{\em
$^1$
Helsinki Institute of Physics, University of Helsinki, PO Box 64, FIN-00014,
Finland \\
$^2$
Department of Physical Science, University of Helsinki, \\PO Box 64, FIN-00014,
Finland \\
$^3$
Department of Physics, Saga University, Saga 840-8502, Japan
}

\end{center}

\vskip .4in

\begin{abstract}

  We discuss the signatures of non-Gaussianity in the curvaton model
  where the potential includes also a non-quadratic term.  In such a
  case the non-linearity parameter $f_{\rm NL}$ can become very small,
  and we show that non-Gaussianity is then encoded in the
  non-reducible non-linearity parameter $g_{\rm NL}$ of the
  trispectrum, which can be very large.  Thus the place to look for
  the non-Gaussianity in the curvaton model may be the trispectrum
  rather than the bispectrum.  We also show that $g_{\rm NL}$ measures
  directly the deviation of the curvaton potential from the purely
  quadratic form.  While $g_{\rm NL}$ depends on the strength of the
  non-quadratic terms relative to the quadratic one, we find that for
  reasonable cases roughly $g_{\rm NL}\sim{\cal O}(-10^4)-{\cal
    O}(-10^5)$, which are values that may well be accessible by future
  observations.

\end{abstract}
\end{titlepage}

\renewcommand{\thepage}{\arabic{page}}
\setcounter{page}{1}
\renewcommand{\thefootnote}{\#\arabic{footnote}}

%%%%%%%%%%%%%%%%%%%%%%%%%%%%
\section{Introduction}
%%%%%%%%%%%%%%%%%%%%%%%%%%%%

The high precision of cosmological observations such as WMAP
\cite{Komatsu:2008hk,Dunkley:2008ie} and the forthcoming Planck
Surveyor Mission \cite{Planck} will soon make it possible to probe the
actual physics of the primordial perturbation rather than merely
describing it. As is well known, the primordial scalar and tensor
power spectra, characterized by spectral indices the and
tensor-to-scalar ratio, can be used to test both models of inflation
and other mechanisms for the generation of the primordial perturbation
such as the curvaton \cite{Enqvist:2001zp,Lyth:2001nq,Moroi:2001ct} or
the modulated reheating scenario
\cite{Dvali:2003em,Kofman:2003nx}. However, many models can imprint
similar features on the primordial power spectrum. Therefore the
possible non-Gaussianity of the photon temperature fluctuation, which
may provide invaluable implications on the physics of the early
universe, has been the focus of much attention recently.

The simplest inflation models generate an almost Gaussian
fluctuation. In contrast, in the curvaton scenario there can arise a
large non-Gaussianity
\cite{Lyth:2002my,Bartolo:2003jx,Enqvist:2005pg,Malik:2006pm,Sasaki:2006kq,Huang:2008ze,Ichikawa:2008iq}.
However, in most studies on the curvaton so far, one simply assumes a
quadratic curvaton potential. Since the curvaton cannot be completely
non-interacting (it has to decay), it is of interest to consider the
implications of the deviations from the exactly quadratic potential,
which represent curvaton self-interactions. Such self-interactions
would arise e.g. in curvaton models based on the flat directions of
the minimally supersymmetric standard model (MSSM)
\cite{MSSMcurvaton}. Even small deviations could be important for
phenomenology, as was pointed out in \cite{Enqvist:2005pg} where it
was shown that the non-Gaussianity predicted by the curvaton model can
be sensitive to the shape of the potential. In particular, the
nonlinearity parameter $f_{\rm NL}$ which quantifies the bispectrum of
primordial fluctuation can, in contrast to the case of the quadratic
curvaton potential, be very small in some cases.

However, the signatures of non-Gaussianity can be probed not only with
the bispectrum but also with the trispectrum.  Thus even if $f_{\rm
  NL}$ is very small, it does not necessarily imply that the
fluctuation is almost Gaussian; rather, in that case the imprint of
non-Gaussianity may be detected only in higher order statistics. The
situation is then more complicated since trispectrum is characterized
by two numbers, conventionally denoted as $\tau_{\rm NL}$ and $g_{\rm
  NL}$.  Roughly, $\tau_{\rm NL}\sim f_{\rm NL}^2$ for single field
models including the curvaton, whereas $g_{\rm NL}$ measures the
four-point correlator that is not reducible to the three-point
correlator. In general, one can expect that when $f_{\rm NL}$ is
large, also $g_{\rm NL}$ is large, although this is not a necessity,
as will be discussed below.

The main purpose of this paper is to investigate non-Gaussianity in
the curvaton model by considering both the bispectrum and the
trispectrum assuming a potential which includes also a non-quadratic
term.  We will show that even when $f_{\rm NL}$ is negligibly small,
$g_{\rm NL}$ can be very large, indicating that the first signature of
the curvaton-induced primordial non-Gaussianity may not come from the
bispectrum but rather from the trispectrum.

The structure of the paper is as follows. In section 2, we summarize
the formalism, describe the set-up of the curvaton model, and give
some key formulae for the study of non-Gaussianity in the curvaton
model.  Then in section \ref{sec:sig}, we investigate non-Gaussianity
in curvaton models, presenting the predictions for the nonlinearity
parameters $f_{\rm NL}$ and $g_{\rm NL}$. We also discuss how the
predictions are affected by the size of the coupling and the power of
a non-quadratic term.  The final section is devoted to the
conclusions.

%%%%%%%%%%%%%%%%%%%%%%%%%%%%
\section{Non-linearity parameters}\label{sec:NL}
%%%%%%%%%%%%%%%%%%%%%%%%%%%%

Our starting point is the $\delta N$ formalism
\cite{Starobinsky:1986fxa,Sasaki:1995aw,Sasaki:1998ug,Lyth:2004gb}
that can be used to calculate the primordial curvature perturbation
$\zeta$. In this approach, the final amplitude of the primordial
curvature perturbation on the uniform energy density hypersurface is
given by $\delta N (t, \vec{x})$ where $\delta N$ is the perturbation
of the number of $e$-folds on a uniform energy density hypersurface
during the final radiation dominated epoch\footnote{
In the curvaton scenario, there may arise two radiation dominated epochs.
The final radiation epoch here means the one after the curvaton decay.
} with respect to the the initial flat time-slice at horizon crossing
during inflation.  In the following, we are interested in the
fluctuations generated from the curvaton field; thus we can write
$\zeta$ as
\begin{equation}
\label{eq:zeta2}
\zeta
=
N_{\sigma} \delta \sigma_\ast
+
\frac{1}{2} N_{\sigma \sigma} (\delta \sigma_\ast )^2
+
\frac{1}{6} N_{\sigma \sigma \sigma} (\delta \sigma_\ast )^3
+ \cdots
\end{equation}
where $\sigma$ is the curvaton field and the derivative of $N$ with
respect to $\sigma$ are represented as $N_\sigma \equiv dN / d\sigma,
N_{\sigma\sigma} \equiv d^2 N/d \sigma^2$ and $N_{\sigma\sigma\sigma}
\equiv d^3 N/d \sigma^3$.

To discuss non-Gaussianity in the scenario, we make use of the
non-linearity parameters $f_{\rm NL}$ and $g_{\rm NL}$ defined by the
expansion
\begin{equation}
\label{eq:defs}
\zeta = \zeta_1 + \frac{3}{5} f_{\rm NL} \zeta_1^2 + \frac{9}{25} g_{\rm NL}\zeta_1^3
+ \mathcal{O}(\zeta_1^4).
\end{equation}
Writing the power spectrum as
\begin{equation}
\label{eq:power}
\langle \zeta_{\vec k_1} \zeta_{\vec k_2} \rangle
=
{(2\pi)}^3 P_\zeta (k_1) \delta ({\vec k_1}+{\vec k_2}),
\end{equation}
the bispectrum and trispectrum are given by
\begin{eqnarray}
\langle \zeta_{\vec k_1} \zeta_{\vec k_2} \zeta_{\vec k_3} \rangle
&=&
{(2\pi)}^3 B_\zeta (k_1,k_2,k_3) \delta ({\vec k_1}+{\vec k_2}+{\vec k_3}).
\label{eq:bi}
\end{eqnarray}
\begin{eqnarray}
\langle
\zeta_{\vec k_1} \zeta_{\vec k_2} \zeta_{\vec k_3} \zeta_{\vec k_4}
\rangle
&=&
{(2\pi)}^3 T_\zeta (k_1,k_2,k_3,k_4) \delta ({\vec k_1}+{\vec k_2}+{\vec k_3}+{\vec k_4}),
\label{eq:tri}
\end{eqnarray}
where $B_\zeta$ and $T_\zeta$ are products of the power spectra and
can be written as
\begin{eqnarray}
B_\zeta (k_1,k_2,k_3)
&=&
\frac{6}{5} f_{\rm NL}
\left(
P_\zeta (k_1) P_\zeta (k_2)
+ P_\zeta (k_2) P_\zeta (k_3)
+ P_\zeta (k_3) P_\zeta (k_1)
\right), \\
\label{eq:def_f_NL}
T_\zeta (k_1,k_2,k_3,k_4)
&=&
\tau_{\rm NL} \left(
P_\zeta(k_{13}) P_\zeta (k_3) P_\zeta (k_4)+11~{\rm perms.}
\right) \nonumber \\
&&
+ \frac{54}{25} g_{\rm NL} \left( P_\zeta (k_2) P_\zeta (k_3) P_\zeta (k_4)
+3~{\rm perms.} \right).
\label{eq:def_tau_g_NL}
\end{eqnarray}
Note the appearance of the independent non-linearity parameter
$\tau_{\rm NL}$.  However, in the scenario we are considering in the
following, $\tau_{\rm NL}$ is related to $f_{\rm NL}$ by
\begin{equation}
\label{eq:taurelated}
\tau_{\rm NL} = \frac{36}{25} f_{\rm NL}^2~.
\end{equation}

To evaluate the primordial curvature fluctuation in the curvaton
model, we need to specify a potential for the curvaton. Here we go
beyond the usual quadratic approximation and consider the following
potential:
\begin{equation}
\label{eq:V}
V(\sigma)
=
\frac{1}{2} m_\sigma^2 \sigma^2
+
\lambda m_\sigma^4 \left( \frac{\sigma}{m_\sigma} \right)^n~,
\end{equation}
which contains a higher polynomial term in addition to the quadratic
term.  For later discussion, we define a parameter $s$ which
represents the size of the non-quadratic term relative to the
quadratic one:
\begin{equation}
\label{eq:def_s}
s \equiv 2 \lambda \left( \frac{\sigma_\ast}{m_\sigma} \right)^{n-2}.
\end{equation}
Thus the larger $s$ is, the larger is the contribution from the
non-quadratic term.

When the potential is quadratic, the fluctuation evolves exactly as
the homogeneous mode. However, when the curvaton field evolves under a
non-quadratic potential, the fluctuation of the curvaton evolves
non-linearly on large scales. In that case the curvature fluctuation
can be written, up to the third order, as \cite{Sasaki:2006kq}
\begin{eqnarray}
%\label{ }
\zeta = \delta  N &=&
\frac{2}{3} r \frac{\sigma'_{\rm osc}}{\sigma_{\rm osc}}  \delta \sigma_\ast
+
\frac{1}{9} \left[ 3r\left(
1 +
\frac{\sigma_{\rm osc} \sigma_{\rm osc}^{\prime\prime}}{\sigma_{\rm osc}^{\prime 2}}
\right)
- 4 r^2 -2  r^3
\right]
\left( \frac{\sigma'_{\rm osc}}{\sigma_{\rm osc}} \right)^2  (\delta \sigma_\ast )^2 \notag \\
&&+
\frac{4}{81} \left[
\frac{9r}{4}  \left(
\frac{\sigma_{\rm osc}^2 \sigma_{\rm osc}^{\prime\prime\prime}}
{\sigma_{\rm osc}^{\prime 3}}
+
3\frac{\sigma_{\rm osc} \sigma_{\rm osc}^{\prime\prime}}{\sigma_{\rm osc}^{\prime 2}}
\right)
-9r^2
\left(
1
+
\frac{\sigma_{\rm osc} \sigma_{\rm osc}^{\prime\prime}}{\sigma_{\rm osc}^{\prime 2}}
\right)
\right. \notag \\
&&
\left.
+\frac{r^3}{2} \left(
1
-
9\frac{\sigma_{\rm osc} \sigma_{\rm osc}^{\prime\prime}}{\sigma_{\rm osc}^{\prime 2}}
\right)
+10r^4 + 3r^5
\right]
\left( \frac{\sigma'_{\rm osc}}{\sigma_{\rm osc}} \right)^3  (\delta \sigma_\ast )^3~,
\end{eqnarray}
where $\sigma_{\rm osc}$ is the value of the curvaton at the onset of
its oscillation while $r$ roughly represents the ratio of the energy
density of the curvaton to the total density at the time of the
curvaton decay.  The exact definition is given by
\begin{equation}
\label{eq:def_r}
\left.
r \equiv \frac{3 \rho_\sigma}{4 \rho_{\rm rad} + 3\rho_\sigma}\right|_{\rm decay}.
\end{equation}
Notice that $\sigma^\prime_{\rm osc}/ \sigma_{\rm osc}=1/\sigma_\ast$
for the case of the quadratic potential.  With this expression, we can
write down the non-linearity parameter $f_{\rm NL}$ as
\begin{equation}
\label{eq:fNL}
f_{\rm NL} = \frac{5}{4r} \left(
1 +
\frac{\sigma_{\rm osc} \sigma_{\rm osc}^{\prime\prime}}{\sigma_{\rm osc}^{\prime 2}}
\right)
- \frac{5}{3} -\frac{5r}{6}.
\end{equation}
Also notice that, although the curvaton scenario generally generates
large non-Gaussianity with $f_{\rm NL} \gtrsim \mathcal{O}(1)$, the
non-linearity parameter $f_{\rm NL}$ can be very small in the presence
of the non-linear evolution of the curvaton field which can render the
term $1 + ( \sigma_{\rm osc} \sigma_{\rm osc}^{\prime\prime} )
/\sigma_{\rm osc}^{\prime 2} \simeq 0$~\cite{Enqvist:2005pg,Sasaki:2006kq}.

In the curvaton model, the non-linearity parameter $g_{\rm NL}$ can be
written as
\begin{equation}
\label{eq:gNL}
g_{\rm NL} = \frac{25}{54}
\left[
\frac{9}{4r^2}  \left(
\frac{\sigma_{\rm osc}^2 \sigma_{\rm osc}^{\prime\prime\prime}}
{\sigma_{\rm osc}^{\prime 3}}
+
3\frac{\sigma_{\rm osc} \sigma_{\rm osc}^{\prime\prime}}{\sigma_{\rm osc}^{\prime 2}}
\right)
-\frac{9}{r}
\left(
1
+
\frac{\sigma_{\rm osc} \sigma_{\rm osc}^{\prime\prime}}{\sigma_{\rm osc}^{\prime 2}}
\right)
+\frac{1}{2} \left(
1
-
9\frac{\sigma_{\rm osc} \sigma_{\rm osc}^{\prime\prime}}{\sigma_{\rm osc}^{\prime 2}}
\right)
+10r + 3r^2
\right].
\end{equation}
As one can easily see, even if the non-linear evolution of $\sigma$
cancels to give a very small $f_{\rm NL}$, such a cancellation does not
necessarily occur for $g_{\rm NL}$.  This indicates an interesting
possibility where the non-Gaussian signature of the curvaton may come
from the trispectrum rather than from the bispectrum\footnote{
  Some general discussions on this issue can be found in
  Ref.~\cite{Sasaki:2006kq}.
}. In the next section, we discuss the ramifications of this
possibility in detail.

%%%%%%%%%%%%%%%%%%%%%%%%%%%%
\section{Signatures of non-Gaussianity}\label{sec:sig}
%%%%%%%%%%%%%%%%%%%%%%%%%%%%

\begin{figure}[t]
\begin{center}
\resizebox{75mm}{!}{\includegraphics{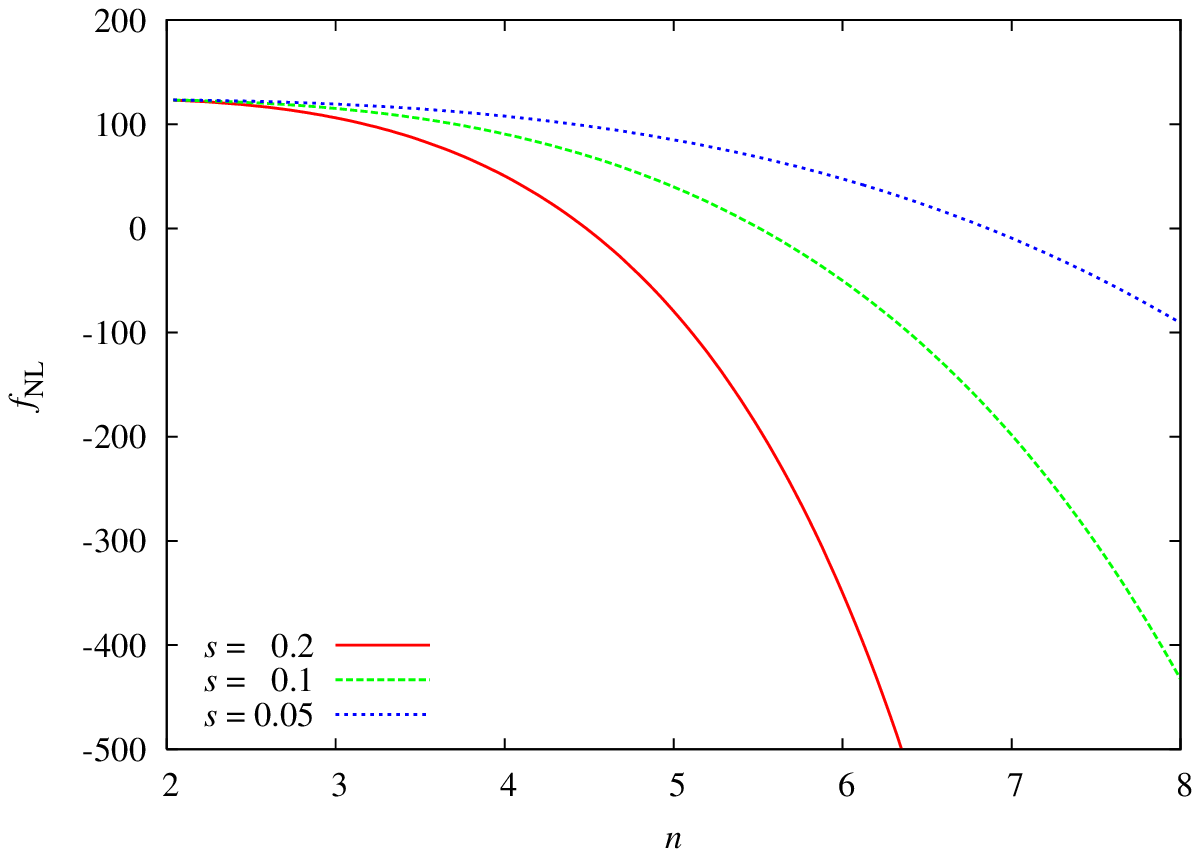}}
\resizebox{75mm}{!}{\includegraphics{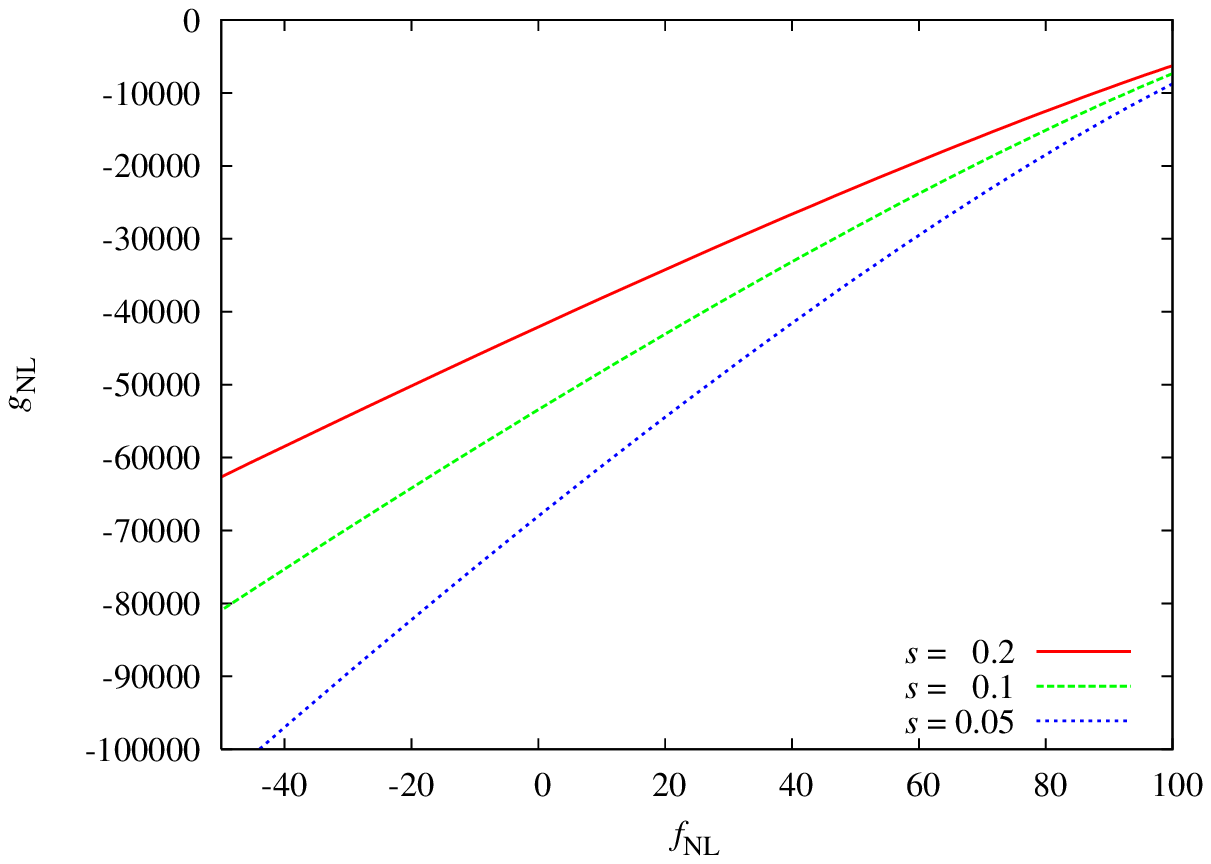}}
\caption{(Left) Plot of $f_{\rm NL}$ as a function of $n$ for several
  values of $s$.  (Right) Plot of $g_{\rm NL}$ as a function of
  $f_{\rm NL}$ for several values of $s$.  Notice that $f_{\rm NL}$
  and $n$ have one-to-one correspondence.  In both panels, $r=0.01$.  }
\label{fig:fNL_gNL_n}
\end{center}
\end{figure}

\begin{figure}[t]
\begin{center}
\resizebox{75mm}{!}{\includegraphics{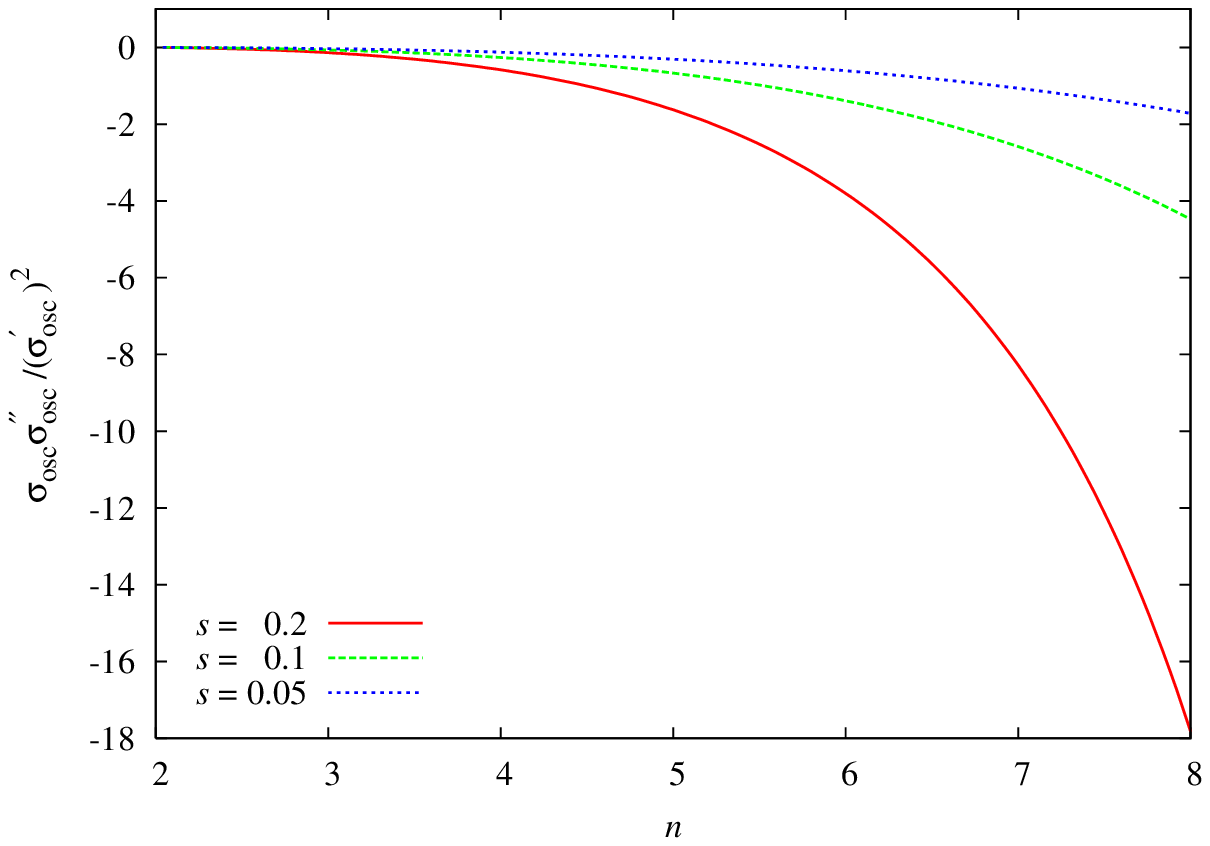}}
\resizebox{75mm}{!}{\includegraphics{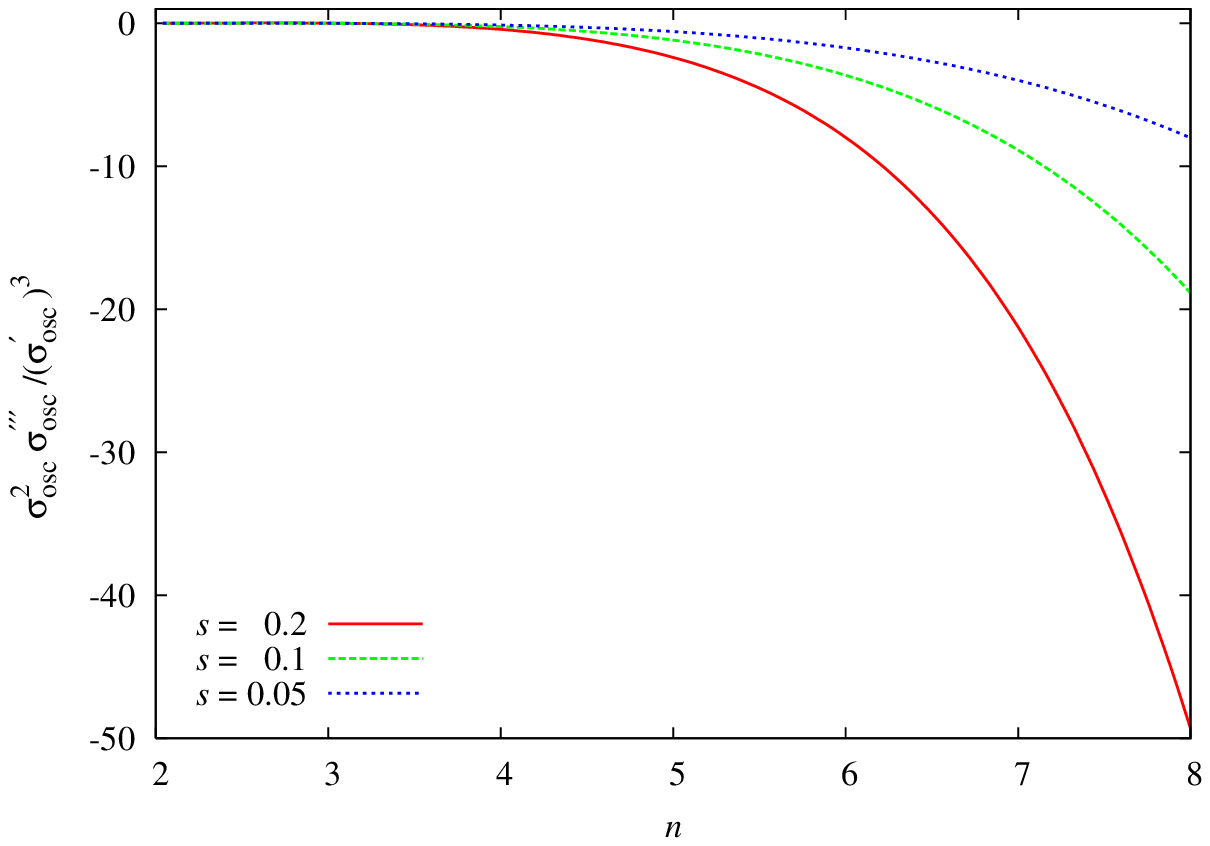}}
\caption{(Left) Plot of the combination $\sigma_{\rm
    osc}\sigma^{\prime\prime}_{\rm osc}/\sigma^{\prime 2}_{\rm osc}$
  as a function of $n$ for several values of $s$.  (Right) Plot of the
  combination $\sigma_{\rm osc}^2\sigma^{\prime\prime\prime}_{\rm
    osc}/\sigma^{\prime 3}_{\rm osc}$ as a function of $n$ for several
  values of $s$. }
\label{fig:combination}
\end{center}
\end{figure}

\begin{figure}[t]
\begin{center}
\resizebox{120mm}{!}{\includegraphics{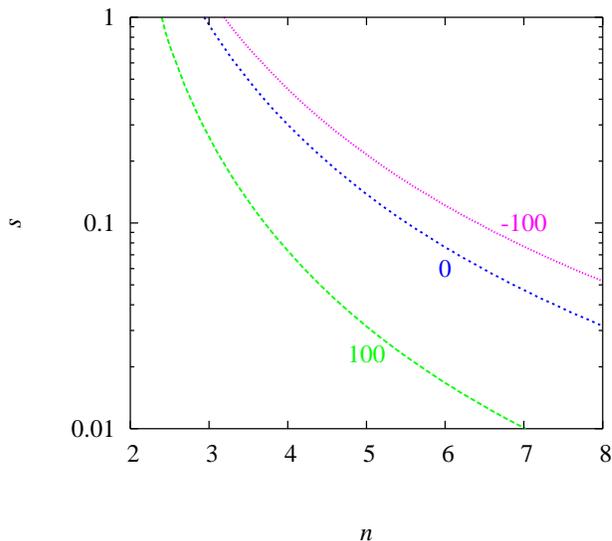}}\vspace{-1cm}
\caption{Contours of $f_{\rm NL}$ are shown for the case $r=0.01$ in
  the $n$--$s$ plane. Note that the line of $f_{\rm NL}=0$ is not
  changed for small values of $r$.}
\label{fig:fNLcont}
\end{center}
\end{figure}

\begin{figure}[t]
\begin{center}
\resizebox{160mm}{!}{\includegraphics{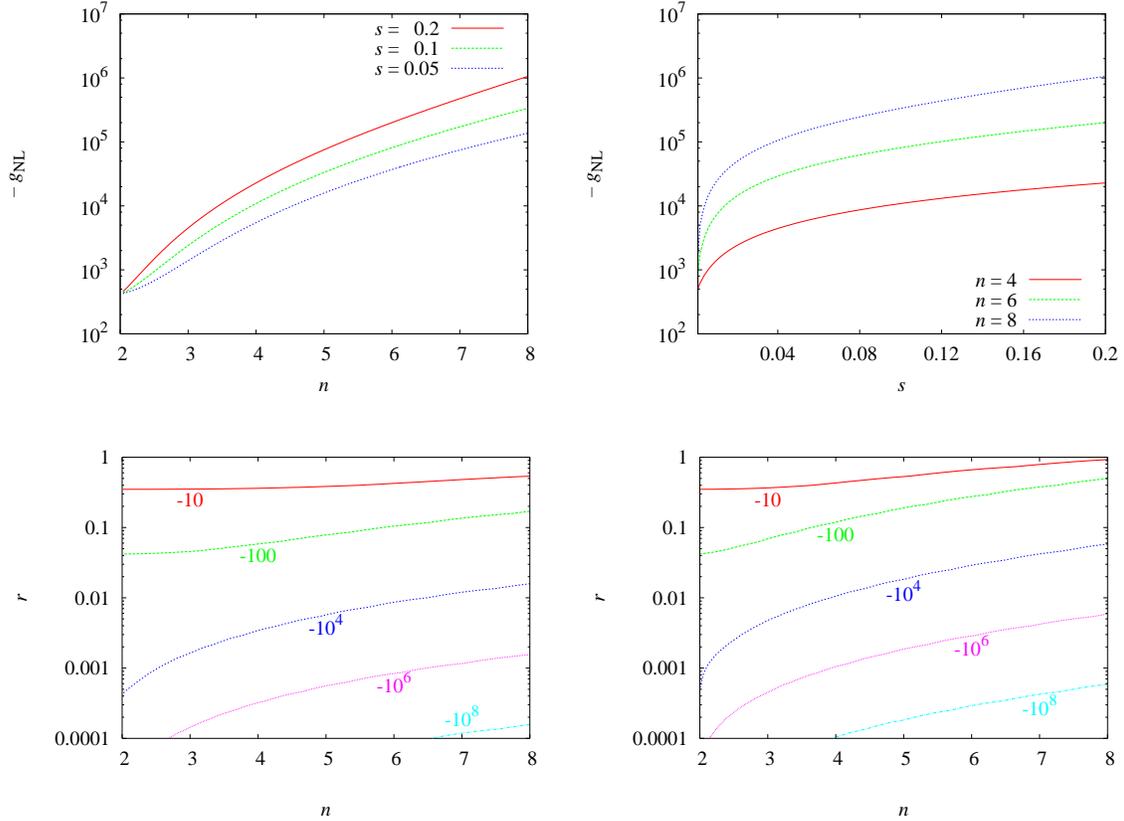}}\vspace{-1cm}
\caption{(Top left) Plot of $-g_{\rm NL}$ as a function of the power
  $n$ for several values of $s$. The value of $r$ is fixed as $r=0.01$
  in the top panels. (Top right) Plot of $-g_{\rm NL}$ as a function of
  $s$ for several values of the power $n$.  (Bottom left) Contours of
  $g_{\rm NL}$ in the $n$--$r$ plane for the case $s=0.1$.  (Bottom
  right) Contours of $g_{\rm NL}$ in the $n$--$r$ plane for the case
  $s=0.01$.}
\label{fig:gNL}
\end{center}
\end{figure}

Let us now consider non-Gaussianity in curvaton models, paying
particular attention to the non-linearity parameters $f_{\rm NL}$ and
$g_{\rm NL}$. We have studied their behavior numerically and plotted
the results in Figs. \ref{fig:fNL_gNL_n}--\ref{fig:gNL}.

In the left panel of Fig.~\ref{fig:fNL_gNL_n}, the value of $f_{\rm
  NL}$ is plotted as a function of the power $n$ for several values of
$s$.  There we have fixed the value of $r$ to $r=0.01$. As can be read
off from Eq.~\eqref{eq:fNL}, when $r$ is small, $f_{\rm NL}$ can be
well approximated as
\begin{equation}
\label{eq:fNL_smallr}
f_{\rm NL} \simeq \frac{5}{4r} \left(
1 +
\frac{\sigma_{\rm osc} \sigma_{\rm osc}^{\prime\prime}}{\sigma_{\rm osc}^{\prime 2}}
\right)~.
\end{equation}
Notice that the combination $\sigma_{\rm osc}
\sigma^{\prime\prime}_{\rm osc} / \sigma_{\rm osc}^{\prime 2}$ is zero
when $n=2$.  As the value of $n$ becomes larger, the above combination
yields a negative contribution, which is depicted in the left panel of
Fig.~\ref{fig:combination}. Thus at some point, $( \sigma_{\rm osc}
\sigma_{\rm osc}^{\prime\prime} ) /\sigma_{\rm osc}^{\prime 2}$
becomes $\sim -1$ and gives $f_{\rm NL}\sim 0$.  In other words,
$f_{\rm NL}$ decreases to zero as the potential deviates away from a
quadratic form and then becomes zero for some values of $n$ and $s$
(see also \cite{Enqvist:2005pg}).  To see in what cases we obtain
$f_{\rm NL}=0$, we show the contours for $f_{\rm NL}$ in
Fig.~\ref{fig:fNLcont}. As seen from the figure, for small values of
$s$, which correspond to the cases where the size of the non-quadratic
term is relatively small compared to the quadratic one, the power $n$
should be large to make $f_{\rm NL}$ very small.  It should also be
mentioned that, for a fixed $s$, if we take larger values of $n$
beyond the ``cancellation point" $f_{\rm NL}=0$, $f_{\rm NL}$ becomes
negative.

However, as already discussed, even if we obtain very small values for
$f_{\rm NL}$, it does not necessarily indicate that non-Gaussianity is
small in the model but may show up in the higher order statistics.
Indeed, this appears to be a generic feature of the curvaton model:
the trispectrum cannot be suppressed and $g_{\rm NL}$ can be quite
large and is always negative for small values of $r$.  This can be
seen in the right panel of Fig.~\ref{fig:fNL_gNL_n}, where we plot the
value of $g_{\rm NL}$ as a function of $f_{\rm NL}$.  Note that once
we choose the value of $n$, $f_{\rm NL}$ is given for fixed $s$ and
$r$. Thus a function of $f_{\rm NL}$ can be regarded as a function of
$n$.  The corresponding value of $n$ for $f_{\rm NL}$ can be read off
from the left panel of Fig.~\ref{fig:fNL_gNL_n}. (We also show the
plot of $g_{\rm NL}$ as a function of $n$ in the top left panel of
Fig.~\ref{fig:gNL}.)

Interestingly, even if $f_{\rm NL}$ is zero, the value of $| g_{\rm
  NL}|$ can be very large, as can be seen Fig.~\ref{fig:fNL_gNL_n}.
This is because a cancellation which can occur for $f_{\rm NL}$ does
not take place for $g_{\rm NL}$, which is a smooth function of
$n$. When $r$ is small, $g_{\rm NL}$ is mainly determined by the first
term in Eq.~\eqref{eq:gNL}:
\begin{equation}
\label{eq:gNL_approx}
g_{\rm NL} \simeq \frac{25}{54}
\left[
\frac{9}{4r^2}  \left(
\frac{\sigma_{\rm osc}^2 \sigma_{\rm osc}^{\prime\prime\prime}}
{\sigma_{\rm osc}^{\prime 3}}
+
3\frac{\sigma_{\rm osc} \sigma_{\rm osc}^{\prime\prime}}{\sigma_{\rm osc}^{\prime 2}}
\right)
\right].
\end{equation}
In the right panel of Fig.~\ref{fig:combination}, we plot the
combination of $( \sigma_{\rm osc}^2 \sigma_{\rm
  osc}^{\prime\prime\prime})/ \sigma_{\rm osc}^{\prime 3}$ as a
function of $n$ for several values of $s$.  There one sees that both
terms in Eq.~\eqref{eq:gNL_approx} give negative contributions, which
indicates that the cancellation between these terms never occurs.
Therefore $g_{\rm NL}$ is always negative for small values of $r$ even
when $f_{\rm NL}$ is very small.  Thus, it may turn out that the best
place to look for non-Gaussianity in curvaton models is the
trispectrum and $g_{\rm NL}$ in particular.  Moreover, since for the
quadratic potential with small $r$, $f_{\rm NL} \simeq 5/4r$ and
$g_{\rm NL} \simeq -10/3r$, in the purely quadratic case there is a
relation between these parameters given by
\begin{equation}
\label{eq:gfrelation}
g_{\rm NL} \simeq -\frac{10}{3} f_{\rm NL}~.
\end{equation}
Thus any deviation from the relation Eq.~(\ref{eq:gfrelation}) can
indicate that the potential is not quadratic.  Therefore the
non-linearity parameter $g_{\rm NL}$ can be a direct measure for the
deviation of the curvaton potential from the quadratic one and hence
probes directly the underlying physics of the model. The larger the
deviation from the purely quadratic case, the more negative $g_{\rm
  NL}$ is.  Note that the right panel of Fig.~\ref{fig:fNL_gNL_n} is
somewhat confusing in this respect. 
To see how $g_{\rm NL}$ depends on the parameters, 
we show several plots for $g_{\rm NL}$ in Fig.~\ref{fig:gNL}.
In the top left and right panels
of Fig. \ref{fig:gNL} we plot $g_{\rm NL}$ for different values of
$s$ and $n$ for a fixed $r=0.01$, respectively, whereas in the bottom
panel of the figure, contours of $g_{\rm NL}$ in the $n$--$r$ plane
for fixed $s=0.1$ (bottom left) $s=0.01$ (bottom right).

%%%%%%%%%%%%%%%%%%%%%%%%%%%%
\section{Conclusion}\label{sec:conclusion}
%%%%%%%%%%%%%%%%%%%%%%%%%%%%

We have discussed the signatures of non-Gaussianity in the curvaton
model with the potential including a non-quadratic term in addition to
the usual quadratic term.  When the curvaton potential is not purely
quadratic, fluctuations of the curvaton field evolve nonlinearly on
superhorizon scales. This gives rise to predictions for the
bispectrum, characterized by the non-linearity parameter $f_{\rm NL}$,
which can deviate considerably from the quadratic case. In particular,
depending the power and the relative strength of a non-quadratic term,
the value of $f_{\rm NL}$ can become zero, as already was pointed out
in Ref.~\cite{Enqvist:2005pg}.  In this paper, we investigated this
issue by considering also the trispectrum, paying particular attention
to the non-linearity parameter $g_{\rm NL}$ which quantifies the the
non-reducible part of the trispectrum. The second non-linearity
parameter describing the trispectrum, denoted as $\tau_{\rm NL}$, is
proportional to $f_{\rm NL}^2$ and hence is small whenever $f_{\rm
  NL}$ is small. In contrast, by studying $f_{\rm NL}$ and $g_{\rm
  NL}$ simultaneously, we find that even when $f_{\rm NL}$ is
negligibly small, the absolute value of $g_{\rm NL}$ can be very
large. Thus the signature of non-Gaussianity in the curvaton model may
come from the trispectrum rather than from the bispectrum if its
potential deviates from a purely quadratic form, as one would expect
in realistic particle physics models.

Moreover, we have shown that the deviation of $g_{\rm NL}$ from the
relation Eq.~(\ref{eq:gfrelation}), valid for the purely quadratic
case in the small $r$ limit, is a direct measure of the deviation from
a quadratic curvaton potential.  In the small $f_{\rm NL}$ limit,
typical values of $g_{\rm NL}$ depend on the strength of the
non-quadratic terms relative to the quadratic one, denoted as $s$ in
Eq.~(\ref{eq:def_s}), but roughly we find that $g_{\rm NL}\sim{\cal
  O}(-10^4)-{\cal O}(-10^5)$, as can be seen in
Fig.~\ref{fig:fNL_gNL_n} and Figs.  \ref{fig:gNL}. Such values may
well be accessible in future experiments such as the Planck Surveyor
Mission. Hence we may soon be in a position to test curvaton models in
a meaningful way by exploring physics beyond the simple
phenomenological quadratic potential.

\bigskip
\bigskip

\noindent {\bf Acknowledgments:} T.T. would like to thank the Helsinki
Institute of Physics for the hospitality during the visit, where this
work was initiated.  This work is supported in part by the Sumitomo
Foundation (T.T.) and the Grant-in-Aid for Scientific Research from
the Ministry of Education, Science, Sports, and Culture of Japan
No.\,19740145 (T.T.), and the Academy of Finland Finnish-Japanese Core
Programme grant 112420.


\begin{thebibliography}{100}

%\cite{Komatsu:2008hk}
\bibitem{Komatsu:2008hk}
  E.~Komatsu {\it et al.}  [WMAP Collaboration],
  %``Five-Year Wilkinson Microwave Anisotropy Probe (WMAP\altaffilmark 1 )
  %Observations:Cosmological Interpretation,''
  arXiv:0803.0547 [astro-ph].
  %%CITATION = ARXIV:0803.0547;%%

  %\cite{Dunkley:2008ie}
\bibitem{Dunkley:2008ie}
  J.~Dunkley {\it et al.}  [WMAP Collaboration],
  %``Five-Year Wilkinson Microwave Anisotropy Probe (WMAP) Observations:
  %Likelihoods and Parameters from the WMAP data,''
  arXiv:0803.0586 [astro-ph].

\bibitem{Planck}
    [Planck Collaboration],
%``Planck: The scientific programme,''
  arXiv:astro-ph/0604069; \\
  Planck webpage,
  {\tt http://www.rssd.esa.int/index.php?project=planck}
  %%CITATION = ASTRO-PH/0604069;%%

%\bibitem{curvaton}
\bibitem{Enqvist:2001zp}
K.~Enqvist and M.~S.~Sloth,
%``Adiabatic CMB perturbations in pre big bang string cosmology,''
Nucl.\ Phys.\ B {\bf 626}, 395 (2002)
[arXiv:hep-ph/0109214];

\bibitem{Lyth:2001nq}
D.~H.~Lyth and D.~Wands,
%``Generating the curvature perturbation without an inflaton,''
Phys.\ Lett.\ B {\bf 524}, 5 (2002)
[arXiv:hep-ph/0110002];

\bibitem{Moroi:2001ct}
T.~Moroi and T.~Takahashi,
%``Effects of cosmological moduli fields on cosmic microwave background,''
Phys.\ Lett.\ B {\bf 522}, 215 (2001)
[Erratum-ibid.\ B {\bf 539}, 303 (2002)]
[arXiv:hep-ph/0110096].
%%%

%\cite{Dvali:2003em}
\bibitem{Dvali:2003em}
  G.~Dvali, A.~Gruzinov and M.~Zaldarriaga,
  %``A new mechanism for generating density perturbations from inflation,''
  Phys.\ Rev.\  D {\bf 69}, 023505 (2004)
  [arXiv:astro-ph/0303591].
  %%CITATION = PHRVA,D69,023505;%%

%\cite{Kofman:2003nx}
\bibitem{Kofman:2003nx}
  L.~Kofman,
  %``Probing string theory with modulated cosmological fluctuations,''
  arXiv:astro-ph/0303614.
  %%CITATION = ASTRO-PH/0303614;%%

%\cite{Lyth:2002my}
\bibitem{Lyth:2002my}
  D.~H.~Lyth, C.~Ungarelli and D.~Wands,
  %``The primordial density perturbation in the curvaton scenario,''
  Phys.\ Rev.\  D {\bf 67}, 023503 (2003)
  [arXiv:astro-ph/0208055].
  %%CITATION = PHRVA,D67,023503;%%


%\cite{Bartolo:2003jx}
\bibitem{Bartolo:2003jx}
  N.~Bartolo, S.~Matarrese and A.~Riotto,
  %``On non-Gaussianity in the curvaton scenario,''
  Phys.\ Rev.\  D {\bf 69}, 043503 (2004)
  [arXiv:hep-ph/0309033].
  %%CITATION = PHRVA,D69,043503;%%


%\cite{Enqvist:2005pg}
\bibitem{Enqvist:2005pg}
  K.~Enqvist and S.~Nurmi,
  %``Non-gaussianity in curvaton models with nearly quadratic potential,''
  JCAP {\bf 0510}, 013 (2005)
  [arXiv:astro-ph/0508573].
  %%CITATION = JCAPA,0510,013;%%


%\cite{Malik:2006pm}
\bibitem{Malik:2006pm}
  K.~A.~Malik and D.~H.~Lyth,
  %``A numerical study of non-gaussianity in the curvaton scenario,''
  JCAP {\bf 0609}, 008 (2006)
  [arXiv:astro-ph/0604387].
  %%CITATION = JCAPA,0609,008;%%

%\cite{Sasaki:2006kq}
\bibitem{Sasaki:2006kq}
  M.~Sasaki, J.~Valiviita and D.~Wands,
  %``Non-gaussianity of the primordial perturbation in the curvaton model,''
  Phys.\ Rev.\  D {\bf 74}, 103003 (2006)
  [arXiv:astro-ph/0607627].
  %%CITATION = PHRVA,D74,103003;%%

 %\cite{Huang:2008ze}
\bibitem{Huang:2008ze}
  Q.~G.~Huang,
  %``Large Non-Gaussianity Implication for Curvaton Scenario,''
  arXiv:0801.0467 [hep-th].
  %%CITATION = ARXIV:0801.0467;%%


%\cite{Ichikawa:2008iq}
\bibitem{Ichikawa:2008iq}
  K.~Ichikawa, T.~Suyama, T.~Takahashi and M.~Yamaguchi,
  %``Non-Gaussianity, Spectral Index and Tensor Modes in Mixed Inflaton and
  %Curvaton Models,''
  arXiv:0802.4138 [astro-ph].
  %%CITATION = ARXIV:0802.4138;%%


\bibitem{MSSMcurvaton}
  K.~Enqvist, S.~Kasuya and A.~Mazumdar,
  %``Adiabatic density perturbations and matter generation from the MSSM,''
  Phys.\ Rev.\ Lett.\  {\bf 90}, 091302 (2003)
  [arXiv:hep-ph/0211147];\\
  %%CITATION = HEP-PH 0211147;%%
%
%
  M.~Postma,
  %``The curvaton scenario in supersymmetric theories,''
  Phys.\ Rev.\ D {\bf 67}, 063518 (2003)
  [arXiv:hep-ph/0212005];\\
  %%CITATION = HEP-PH 0212005;%%
%
%
  K.~Enqvist, A.~Jokinen, S.~Kasuya and A.~Mazumdar,
  %``MSSM flat direction as a curvaton,''
  Phys.\ Rev.\ D {\bf 68}, 103507 (2003)
  [arXiv:hep-ph/0303165];\\
  %%CITATION = HEP-PH 0303165;%%
%
%
  S.~Kasuya, M.~Kawasaki and F.~Takahashi,
  %``MSSM curvaton in the gauge-mediated SUSY breaking,''
  Phys.\ Lett.\ B {\bf 578}, 259 (2004)
  [arXiv:hep-ph/0305134];\\
  %%CITATION = HEP-PH 0305134;%%
%
%
 K.~Enqvist,
  %``Curvatons in the minimally supersymmetric standard model,''
  Mod.\ Phys.\ Lett.\  A {\bf 19}, 1421 (2004)
  [arXiv:hep-ph/0403273];\\
  %%CITATION = MPLAE,A19,1421;%%
  %
  %
  M.~Ikegami and T.~Moroi,
  %``Curvaton scenario with Affleck-Dine baryogenesis,''
  Phys.\ Rev.\ D {\bf 70}, 083515 (2004)
  [arXiv:hep-ph/0404253];\\
  %%CITATION = HEP-PH 0404253;%%
%
%
  R.~Allahverdi, K.~Enqvist, A.~Jokinen and A.~Mazumdar,
  %``Identifying the curvaton within MSSM,''
  JCAP {\bf 0610} (2006) 007
  [arXiv:hep-ph/0603255].
  %%CITATION = JCAPA,0610,007;%%



%\cite{Starobinsky:1986fxa}
\bibitem{Starobinsky:1986fxa}
  A.~A.~Starobinsky,
  %``Multicomponent de Sitter (Inflationary) Stages and the Generation of
  %Perturbations,''
  JETP Lett.\  {\bf 42} (1985) 152
  [Pisma Zh.\ Eksp.\ Teor.\ Fiz.\  {\bf 42} (1985) 124].
  %%CITATION = ZFPRA,42,124;%%

%\cite{Sasaki:1995aw}
\bibitem{Sasaki:1995aw}
  M.~Sasaki and E.~D.~Stewart,
  %``A General Analytic Formula For The Spectral Index Of The Density
  %Perturbations Produced During Inflation,''
  Prog.\ Theor.\ Phys.\  {\bf 95}, 71 (1996)
  [arXiv:astro-ph/9507001].
  %%CITATION = ASTRO-PH 9507001;%%

%\cite{Sasaki:1998ug}
\bibitem{Sasaki:1998ug}
  M.~Sasaki and T.~Tanaka,
  %``Super-horizon scale dynamics of multi-scalar inflation,''
  Prog.\ Theor.\ Phys.\  {\bf 99}, 763 (1998)
  [arXiv:gr-qc/9801017].
  %%CITATION = GR-QC 9801017;%%

%\cite{Lyth:2004gb}
\bibitem{Lyth:2004gb}
  D.~H.~Lyth, K.~A.~Malik and M.~Sasaki,
  %``A general proof of the conservation of the curvature perturbation,''
  JCAP {\bf 0505}, 004 (2005)
  [arXiv:astro-ph/0411220].
  %%CITATION = JCAPA,0505,004;%%




\end{thebibliography}
\end{document}